\documentclass[aps, pre, twocolumn, a4paper, superscriptaddress]{revtex4}

\usepackage{graphicx}
\usepackage[caption=false]{subfig}
\usepackage{float}
\usepackage{bbm}
\usepackage[all]{xy}
\usepackage{amsmath, amssymb}
\usepackage[T1]{fontenc}
\usepackage[utf8]{inputenc}
\usepackage[english]{babel}
\usepackage{enumitem}
\usepackage[normalem]{ulem}
\usepackage{times}

\begin{document}

\allowdisplaybreaks[3]
\newcommand{\stl}[1]{
  \mbox{$
  	\hspace{0.1em}
  	\stackrel{
			\rule{0.4pt}{0.275ex}
    	\hspace{0.40em}
			\!\!\!
    	\overline{
      	\hspace{0.06em}
      	\vphantom{\rule{0.4pt}{0.0ex}}
      	\hphantom{\mbox{$\displaystyle #1$}}
      	\hspace{0.06em}
    	}
			\!\!\!
			\hspace{0.40em}\rule{0.4pt}{0.275ex}
  	}
		{#1}
		\hspace{0.2em}$
  }
}

\title{Electric field induced inversion of the sign of half-integer disclinations in 2D nematic liquid crystals}

\author{P.P. Avelino}
\affiliation{Centro de F\'{\i}sica do Porto, Rua do Campo Alegre 687,
4169-007 Porto, Portugal}
\affiliation{Departamento de F\'{\i}sica da Faculdade de Ci\^encias da
Universidade do Porto, Rua do Campo Alegre 687, 4169-007 Porto,
Portugal}
\affiliation{Departamento de F\'{\i}sica, Universidade Federal da Para\'{\i}ba,
POB 5008 Jo\~ao Pessoa, PB 58051-970, Brazil}
\author{F. Moraes} 
\affiliation{Departamento de F\'{\i}sica, Universidade Federal da Para\'{\i}ba,
POB 5008 Jo\~ao Pessoa, PB 58051-970, Brazil}
\author{J.C.R.E. Oliveira} 
\affiliation{Centro de F\'{\i}sica do Porto, Rua do Campo Alegre 687,
4169-007 Porto, Portugal}
\affiliation{Departamento de Engenharia F\'{\i}sica da Faculdade de
Engenharia da Universidade do Porto, Rua Dr. Roberto Frias, s/n,
4200-465 Porto, Portugal}
\author{B. F. de Oliveira} 
\affiliation{Centro de F\'{\i}sica do Porto, Rua do Campo Alegre 687,
4169-007 Porto, Portugal}
\affiliation{Departamento de F\'{\i}sica da Faculdade de Ci\^encias da
Universidade do Porto, Rua do Campo Alegre 687, 4169-007 Porto,
Portugal}
\affiliation{Departamento de F\'{\i}sica, Universidade Federal da Para\'{\i}ba,
POB 5008 Jo\~ao Pessoa, PB 58051-970, Brazil}

\begin{abstract}

We study the effect of the rotation of an external electric field on the 
dynamics of half-integer disclination networks in two dimensional 
nematic liquid crystals with a negative dielectric anisotropy using 
LICRA, a LIquid CRystal Algorithm developed by the authors. We 
show that a rotation of $\pi$ of the electric field around an axis of 
the liquid crystal plane continuously transforms all half-integer 
disclinations of the network into disclinations of opposite sign via twist 
disclinations. We also determine the evolution of the characteristic 
length scale, thus quantifying the impact of the external electric 
field on the coarsening of the defect network.

\end{abstract}

\maketitle

\section{INTRODUCTION}
\label{introduction} 

Topological defects play a fundamental role in condensed matter 
physics \cite{deGennes-book,Kleman-book} and cosmology 
\cite{1994csot.book.....V}. Liquid crystals are an example of a very 
rich environment where the dynamics of topological defects 
can be realized experimentally at relatively low costs, making 
them ideal laboratories for testing different scenarios for defect formation 
and evolution \cite{Chuang-Science.251.4999, 
Chuang-PhysRevE.47.3343, Zapotocky-PhysRevE.51.1216, 
Digal-PhysRevLett.83.5030, Denniston-PhysRevE.64.021701, 
Dutta-PhysRevE.71.026119, Lozar-PhysRevE.72.051713, 
Mukai-PhysRevE.75.061704, Bhattacharjee-PhysRevE.78.026707}. 
External fields, such as electric and magnetic fields, can have a 
significant impact on the orientational order of liquid crystals 
\cite{Kitzerow-MolCrystLiqCryst.202.51, Dierking-PhysRevE.71.061709, 
Alexander-EPL.81.66004, Fukuda-PhysRevE.80.031706, 
Fukuda-PhysRevE.81.040701, deOliveira-PhysRevE.82.041707} thus 
affecting the coarsening dynamics of disclination networks.
Recently it was shown that the orientation of the applied electric field 
and the sign of the LC dielectric anisotropy may be used to control the 
type and topological charge of disclination networks 
\cite{deOliveira-PhysRevE.82.041707}. In \cite{Fukuda-PhysRevE.81.040701} a 
continuous transformation of some $-1/2$ wedge disclination lines into 
$1/2$ ones was numerically simulated on a cholesteric blue phase of a chiral liquid crystal,  
through the application of a constant electric field oblique to the 
line wedge. This provided a numerical realization of a continuous transformation 
between half-integer disclinations with topological charge of opposite 
sign \cite{Kleman-book}. 

In the present work we provide a numerical realization of the inversion of 
the sign of all half-integer disclinations on a 2D nematic LC with a negative 
dielectric constant induced by the rotation of an external electric field. In our 
implementation all half-integer disclinations of the network transform into 
disclinations of opposite sign via twist disclinations. We also consider other 
dynamical effects associated with the application of the extern electric field 
on the nematic, by comparing the coarsening of the network under the rotation 
of the electric field with its evolution under a transformation of the director profile 
implemented by hand.

The paper is organized as follows. In Sec. \ref{the_model}, the equations 
describing  the relaxational dynamics of nematic LC, in terms of a symmetric, 
traceless order parameter $Q_{\alpha \beta}$, are presented. The numerical 
techniques are discussed in Sec. \ref{numerical_implementation}. In Sec. 
\ref{results} the effect of the rotation of an external electric field on the 
evolution of half-integer disclination networks on a two-dimensional nematic 
LC with a negative dielectric anisotropy is studied in detail. The conclusions and 
final remarks are presented in Sec. \ref{conclusion}

\section{THE MODEL}
\label{the_model} 

The orientational order of a nematic LC without intrinsic biaxiality is
described by a symmetric traceless tensor, $Q_{\alpha\beta}$, at every
point in space, whose components are given by \cite{deGennes-book}
\begin{equation}
	Q_{\alpha\beta}= \frac{3}{2}S(n_{\alpha}n_{\beta} -
\frac{1}{3}\delta_{\alpha\beta}) + \frac{1}{2}T\left(l_{\alpha}l_{\beta}
- m_{\alpha}m_{\beta}\right)\,, \label{Q_ab}
\end{equation}
where the unit vector ${\bf n}$ is the director, determining the local average 
orientation of the molecules, ${\bf l}$ is the codirector, 
associated with the direction of orientational order
perpendicular to ${\bf n}$ and ${\bf m}={\bf
n}\times{\bf l}$. The variables $S$ and $T$ represent the strength of
uniaxial and biaxial ordering, respectively. The values of $S$ and $T$
may be found by the diagonalization of the matrix
\begin{equation} 
	Q_{\alpha\beta}=\left(
		\begin{array}{ccc} 
			-(S + T)/2     & 0          & 0\\ 
			0              & -(S - T)/2 & 0\\ 
			0              & 0          & S
		\end{array}\right)\,,
	\label{eigenvalues}
\end{equation}
in a coordinate system where ${\bf n}=(0,0,1)$, ${\bf l}=(0,1,0)$ and 
${\bf m}=(1,0,0)$.

Static equilibrium can only be reached for a minimum value of the free
energy ($\delta F/\delta Q_{\alpha\beta}=0$). However, the evolution of
the order parameter $Q_{\alpha\beta}$ from a given set of initial
conditions is not fully specified by the free energy functionals, and
further assumptions have to be made on how the minimization process will
take place. In the absence of thermal fluctuations and hydrodynamic 
flow, the time evolution of the order parameter is given by
\cite{deGennes-MolCrystLiqCryst.12.193}
\begin{equation}
	{\dot{Q}}_{\alpha\beta}({\bf r},t)=
-\Gamma_{\alpha\beta\mu\nu}\frac{\delta F}{\delta Q_{\mu\nu}}\,.
\label{eq_of_motion_1}
\end{equation}
Here the dot represents derivative with respect to the physical time,
$t$, and the tensor
\begin{equation}
	\Gamma_{\alpha\beta\mu\nu}=
\Gamma\left(\delta_{\alpha\mu}\delta_{\beta\nu} +
\delta_{\alpha\nu}\delta_{\beta\mu} -
\frac{2}{3}\delta_{\alpha\beta}\delta_{\mu\nu}\right)\,,
\label{Gamma}
\end{equation}
satisfies $\Gamma_{\alpha\beta\mu\nu}= \Gamma_{\beta\alpha\mu\nu}=
\Gamma_{\mu\nu\alpha\beta}$ and $\Gamma_{\alpha\alpha\mu\nu}=0$ thus
ensuring that the order parameter $Q_{\alpha\beta}$ remains symmetric
and traceless. In the following we shall assume that the kinetic
coefficient, $\Gamma$, is a constant.

\begin{figure}[htb]
	\includegraphics[scale=0.38]{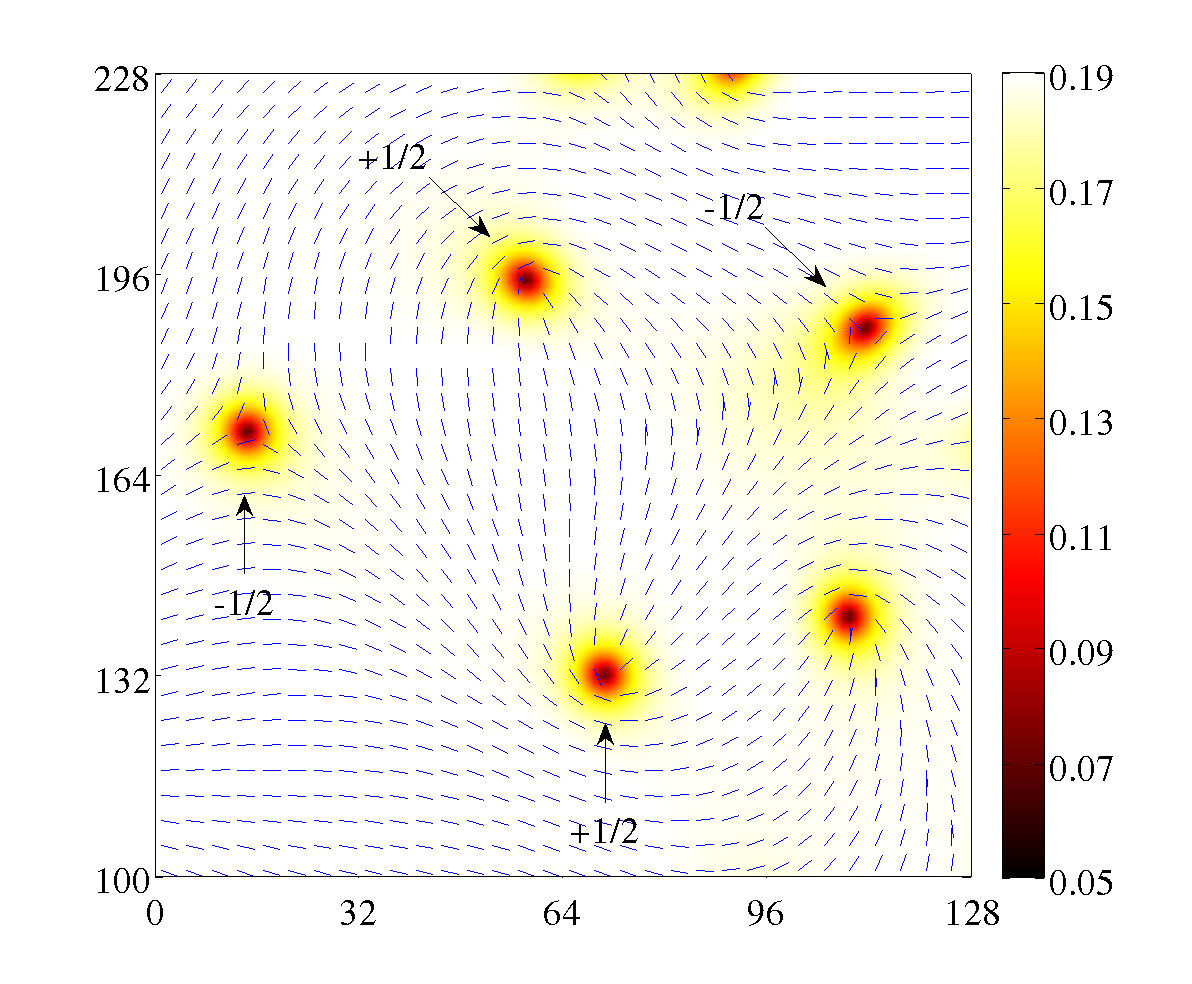} 
	\caption{{\small (Color online) Projection of the director ${\bf n}$
onto the two dimensional grid along with the value of the parameter $S$
at every grid point, at $t= 499$. Only half-integer disclinations remain 
due to application of an external electric field along the $z$ direction.}}
	\label{S_n-500}
\end{figure}

The free energy can be written as
\begin{equation}
	F= \int d^{3}{\bf r} \left(\mathcal{F}_{b} + \mathcal{F}_{el} +
\mathcal{F}_{E}\right)\,.
	\label{free_energy}
\end{equation}
The first term, $\mathcal{F}_{b}$, is the bulk free energy density. It describes the
nematic-isotropic phase transition and it may be obtained form a local
expansion in rotationally invariant powers of the order
parameter
\begin{equation} 
	\mathcal{F}_{b}= \frac{A}{2}\,{\rm Tr}Q^{2} + \frac{B}{3}\,{\rm
Tr}Q^{3} + \frac{C}{4}\,({\rm Tr}Q^{2})^{2}\,.
	\label{energy_bulk}
\end{equation}
The second term, $\mathcal{F}_{el}$, is the elastic free energy density. 
Using the one elastic constant approximation the elastic free energy 
density is given
by
\begin{equation}
	\mathcal{F}_{el}= \frac{R}{2}\, \partial_{\alpha}Q_{\beta\gamma}\,
\partial_{\alpha}Q_{\beta\gamma}\,,
	\label{energy_elastic}
\end{equation}
where the constant $R$ is the single elastic constant.
The last term in the free energy functional, $\mathcal{F}_{E}$, is the
contribution of the effect of an external electric field ${\bf E}$,
\begin{equation}
	\mathcal{F}_{E}= -\frac{\tau_{E}}{2}
Q_{\alpha\beta}E_{\alpha}E_{\beta}\,,
	\label{Electric_fiel}
\end{equation}
where $\tau_{E}= 2\Delta\epsilon$ and $\Delta\epsilon$ is the dielectric
anisotropy. The molecules of a LC with positive dielectric constant,
$\Delta\epsilon>0$, tend to orient parallel to the external electric
field, ${\bf E}$. On the other hand, if the LC has a negative dielectric constant,
$\Delta\epsilon<0$, then the molecules tend to be align perpendicularly 
to ${\bf E}$.

\begin{figure}[htb]
	\includegraphics[scale=0.38]{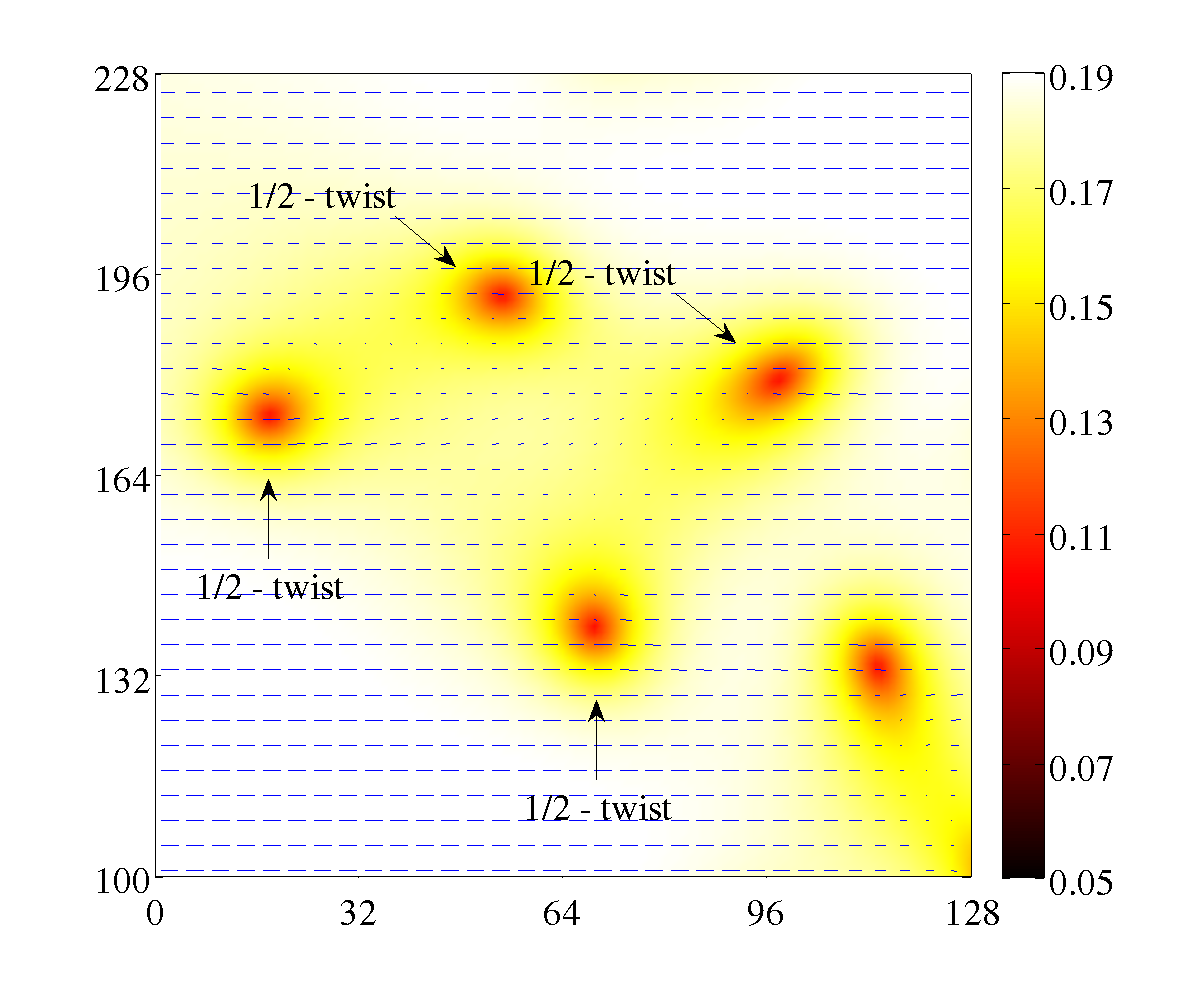} 
	\caption{{\small (Color online) Projection of the director ${\bf n}$
onto the two dimensional grid along with the value of the parameter $S$
at every grid point for a single snapshot of simulation $B$ at $t= 650$, after the rotation of the
electric field by an angle of $\pi/2$. Only half-integer twist disclinations 
appear in this snapshot.}}
	\label{S_n-650-B} 
\end{figure}

With this free energy the equation of motion can be written more
explicitly as
\begin{eqnarray}
	{\dot Q}_{\alpha\beta}  = & - \Gamma \left[ \right. (A + C\, {\rm
Tr}Q^{2})Q_{\alpha\beta} + B\, \stl{Q_{\alpha\gamma}Q_{\gamma\beta}}
\nonumber\\
	&  - R\, Q_{\alpha\beta,\gamma\gamma} -\tau_E \stl{E_\alpha E_\beta}\left. \right]\,,
	\label{eq_of_motion_2}
\end{eqnarray}
where a comma denotes a partial derivative and $\stl{X_{\alpha\beta}}$
setting an arbitrary real matrix, $X_{\alpha\beta}$, to be a symmetric
traceless matrix by the following procedure
\begin{equation}
	\stl{X_{\alpha\beta}}= \frac{1}{2} (X_{\alpha\beta} + X_{\beta\alpha})
- \frac{1}{3} \delta_{\alpha\beta} X_{\gamma\gamma}\,.
	\label{symmetric_traceless}
\end{equation}

\section{NUMERICAL IMPLEMENTATION}
\label{numerical_implementation} 

All the numerical simulations were performed with LICRA (LIquid CRystal Algorithm), a 
publicly available set of C codes and {\small
MATLAB/OCTAVE} routines used to solve Eq. (\ref{eq_of_motion_2}) 
using a standard second-order finite difference algorithm for the spatial
derivatives and a second order Runge-Kutta method for the time
integration. This software is free and it is available at
\textbf{http://faraday.fc.up.pt/licra} \cite{deOliveira-PhysRevE.82.041707}. 

The order parameter ${\bf Q}$ has 5 degrees of freedom associated 
with $S$, $T$, $\bf n$ and $\bf l$ ($\bf n$ accounts for two degrees of 
freedom). The initial conditions for $S$ and $T$ are randomly generated, 
at every grid point, from uniform distributions in the intervals $[0,2/3]$ and 
$[0,3S]$, respectively. The director, ${\bf n}$, is also randomly generated, 
at every grid point, using the spherical vector distribuitions routines in the 
GNU Scientific Library (GSL) and the codirector, $\bf l$ , was calculated by randomly 
choosing a direction perpendicular to ${\bf n}$. The eigenvalues and 
eigenvectors in LICRA are computed using the library GSL.

\begin{figure}[htb]
	\includegraphics[scale=0.38]{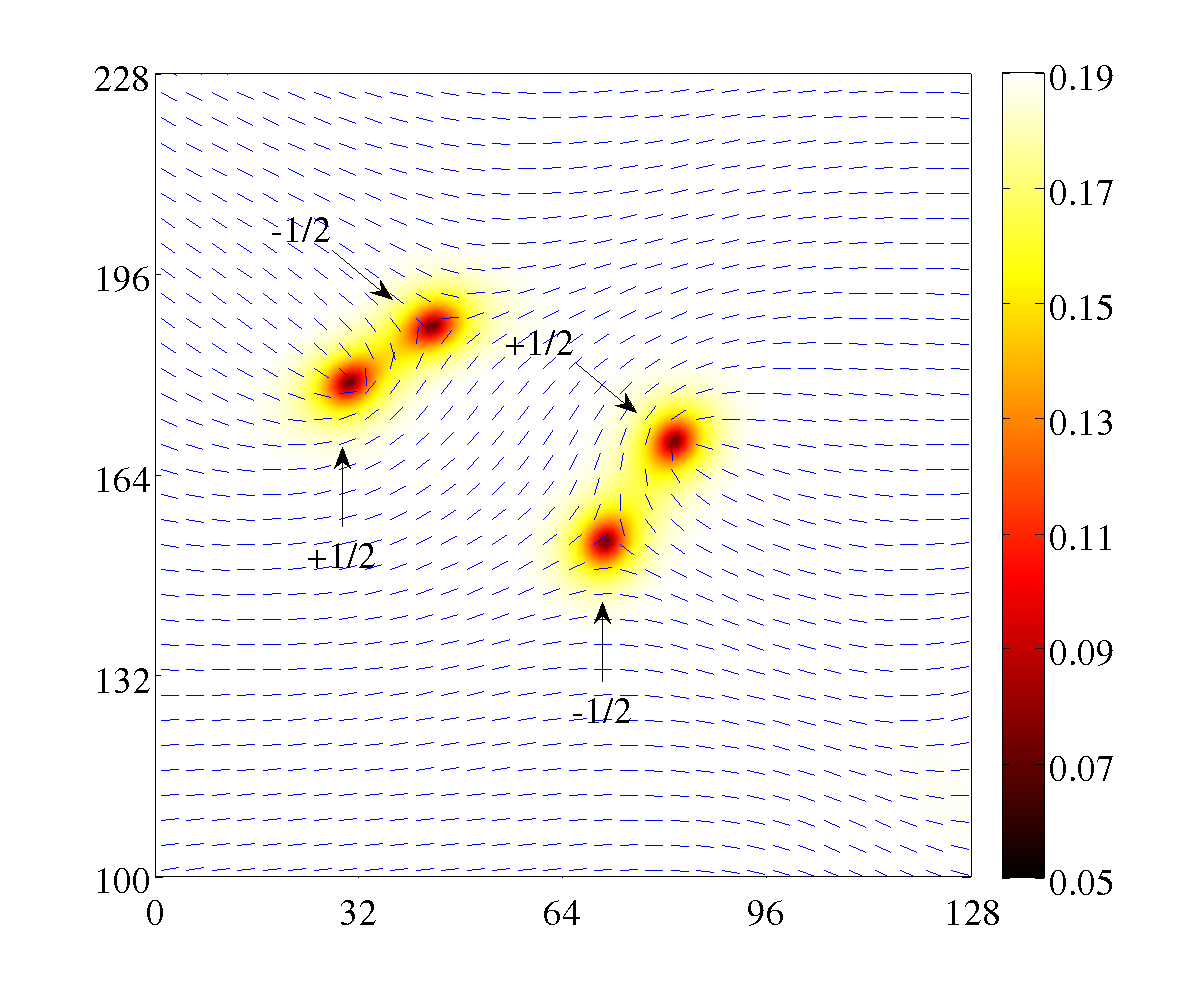}
	\caption{{\small (Color online) Similar to Figs. \ref{S_n-650-B} and 
	\ref{S_n-500} but now at $t= 900$, ($t>t_{\rm off}$). The sign of all
topological charges were inverted with respect to Fig.
\ref{S_n-500}.}}
\label{S_n-900-B} 
\end{figure}

In order to determine the evolution of the characteristic scale of the
network LICRA numerically calculates the correlation function in Fourier
space
\begin{equation}
	P({\bf k},t)= \frac{Q_{\alpha\beta}({\bf k},t) Q_{\beta\alpha}({\bf
-k},t)} {\int{d^{3}{\bf k}\; Q_{\alpha\beta}({\bf k},t)
Q_{\beta\alpha}({\bf -k},t)}}\,,
	\label{function_P}
\end{equation}
after setting the infinite wavelength mode $Q_{\alpha\beta}({\bf 0})$ to
zero. Here $Q_{\alpha\beta}({\bf k})$ is the Fourier transform of
$Q_{\alpha\beta}({\bf r})$ (${\bf k}$ is the wavenumber) and was
calculated by using the library Fastest Fourier Transform in the West.
The characteristic scale, $L$, can then be defined as
\begin{equation}
	\frac{1}{L^{2}}= \langle k^{2}\rangle= \sum_{{\bf k}}k^{2}P({\bf
k},t)/\sum_{{\bf k}}P({\bf k},t)\,.
\label{length_scale}
\end{equation}

In this paper we have used LICRA to perform high resolution numerical 
simulations of the dynamics of a texture network in a uniaxial 
nematic two-dimensional LC under an external electric field.
All simulations were performed on a  $2048^{2}$ grid (in
the $xy$ plane)  with parameters $\Delta x=\Delta y =1$, $\Delta t=0.1$, 
$A= -0.1$, $B= -0.5$, $C=2.67$, $R= 1.0$, $\Gamma = 1.0$, 
$|{\bf E}|=0.2/{\sqrt {|\Delta \epsilon|}}$ and $\Delta \epsilon =-0.04$. 

\section{RESULTS}
\label{results} 

The evolution of $Q_{\alpha\beta}$ starts with the external electric
field switched off. An electric field perpendicular to the plane of the 
simulation (along the $z$ direction) is connected at $t= 300$ 
and disconnected at $t= 350$. This procedure ensures that only
half-integer disclinations remain in the simulation
\cite{deOliveira-PhysRevE.82.041707}. After that, three different 
simulations have been performed: simulation $A$ with no external electric
field, simulation $B$ where a rotating external electric field has been applied 
between $t_{\rm on}= 500$ and $t_{\rm off}= 800$ and simulation 
$C$ where a modification by hand of the director and co-director 
profiles was made at $t= 500$. Fig. \ref{S_n-500} shows the
value of the order parameter $S$ and the projection of the director
${\bf n}$ onto the two dimensional grid for $t=499$, which is equivalent
for simulation $A$, $B$ and $C$.

\begin{figure}[htb]
	\includegraphics[scale=0.38]{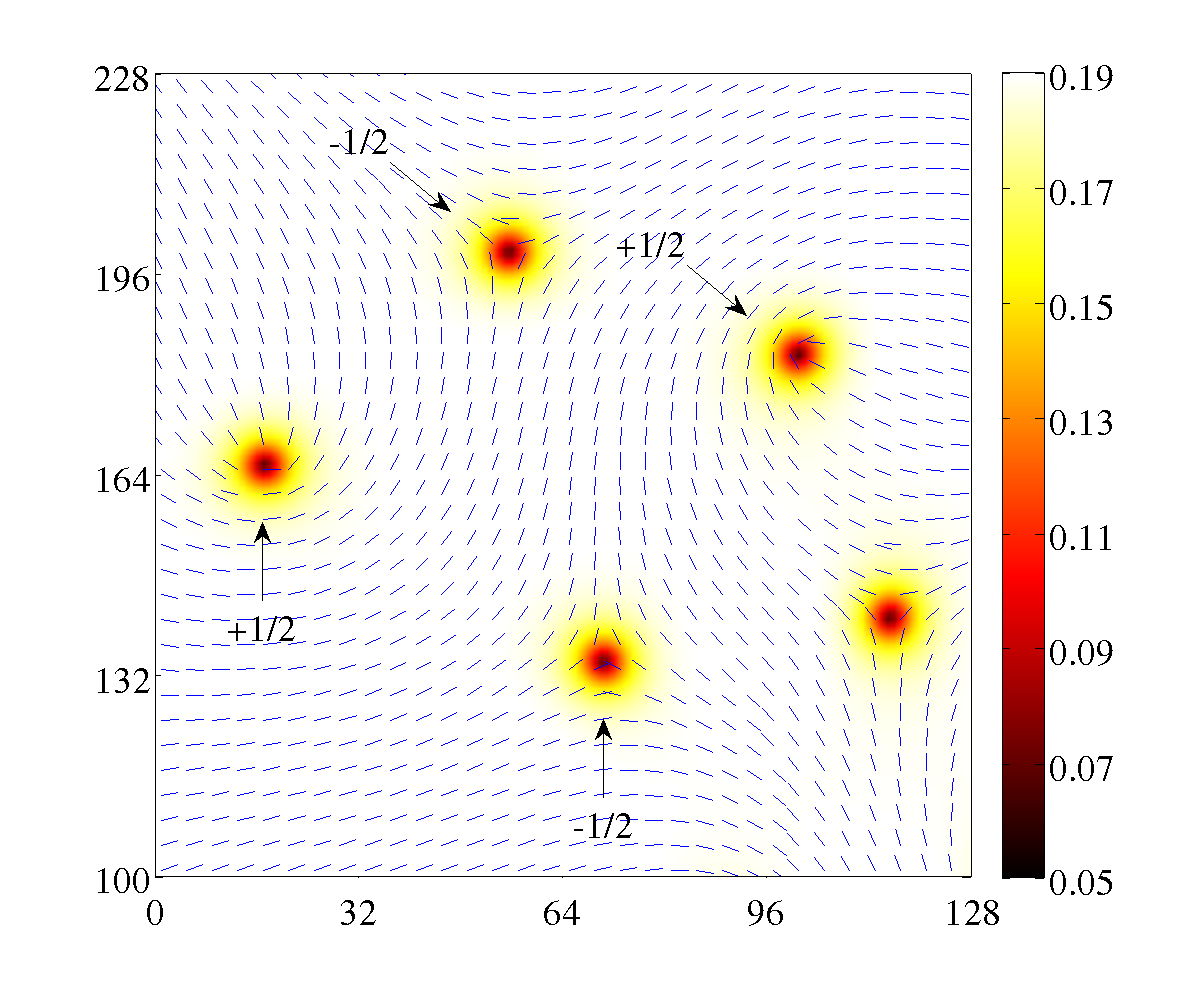}
	\caption{{\small (Color online) Projection of the director ${\bf n}$
onto the two dimensional grid along with the value of the parameter $S$
at every grid point of a single snapshot of simulation $C$ at $t= 900$, after changing the director
configuration profile by hand according to Eq. (\ref{changedir}). The sign of all
topological charges were inverted with respect to Fig. \ref{S_n-500}.}}
\label{S_n-900-C} 
\end{figure}

In simulation $B$ the components of the electric field are given by
\begin{eqnarray} 
	E_{x} & = & 0\,,\nonumber \\ 
	E_{y} & = & |{\bf E}|\sin(\varphi)\,,\nonumber \\ 
	E_{z} & = & |{\bf E}|\cos(\varphi)\,,
\label{components_E}
\end{eqnarray}
with
\begin{equation}
\varphi=\frac{\pi}{300} \left(t-500\right)\,.
\label{length_scale}
\end{equation}
The electric field was applied at $t_{\rm on}= 500$ along the $z$ direction 
and then rotates around the $x$ direction by an angle of $\varphi$, on 
the $yz$ plane. At each time step ($\Delta t=0.1$) the angle, $\varphi$,
was increased by $\pi/3000$ radians until a total angle of $\pi$ 
(at $t_{\rm off}= 800$). The negative dielectric anisotropy, $\Delta\epsilon < 0$, 
ensures that the director remains perpendicular to the field ensuring that ${\bf
n}\cdot{\bf E}= 0$. If the defect centers were static then, at each time-step, the 
components of the director would be equal to
\begin{eqnarray} 
	n_{x} & = &  n_{x_{0}} \,,\nonumber \\ 
	n_{y} & = &  n_{y_{0}}\cos(\varphi) \,,\nonumber \\ 
	n_{z} & = & -n_{y_{0}}\sin(\varphi)\,,
	\label{components_n}
\end{eqnarray}
where $(n_{x_{0}},n_{y_{0}},0)$ are the components of the director
at $t= 500$ in simulation $A$. In practice this result must be complemented with the 
coarsening dynamics of the network.

Fig. \ref{S_n-650-B} presents the order parameter $S$ and the 
projection of the director, ${\bf n}$, onto the two dimensional grid 
for a snapshot taken from simulation $B$, at $t= 650$, when $\varphi=\pi/2$. In this case, the 
director is given by  ${\bf n}=(n_{x_{0}},0,-n_{y_{0}})$ and twist disclinations 
are formed on the $xz$ plane. Note that the presence of the electric field 
increases the size of the defect cores.

\begin{figure}[htb]
	\includegraphics[scale=0.38]{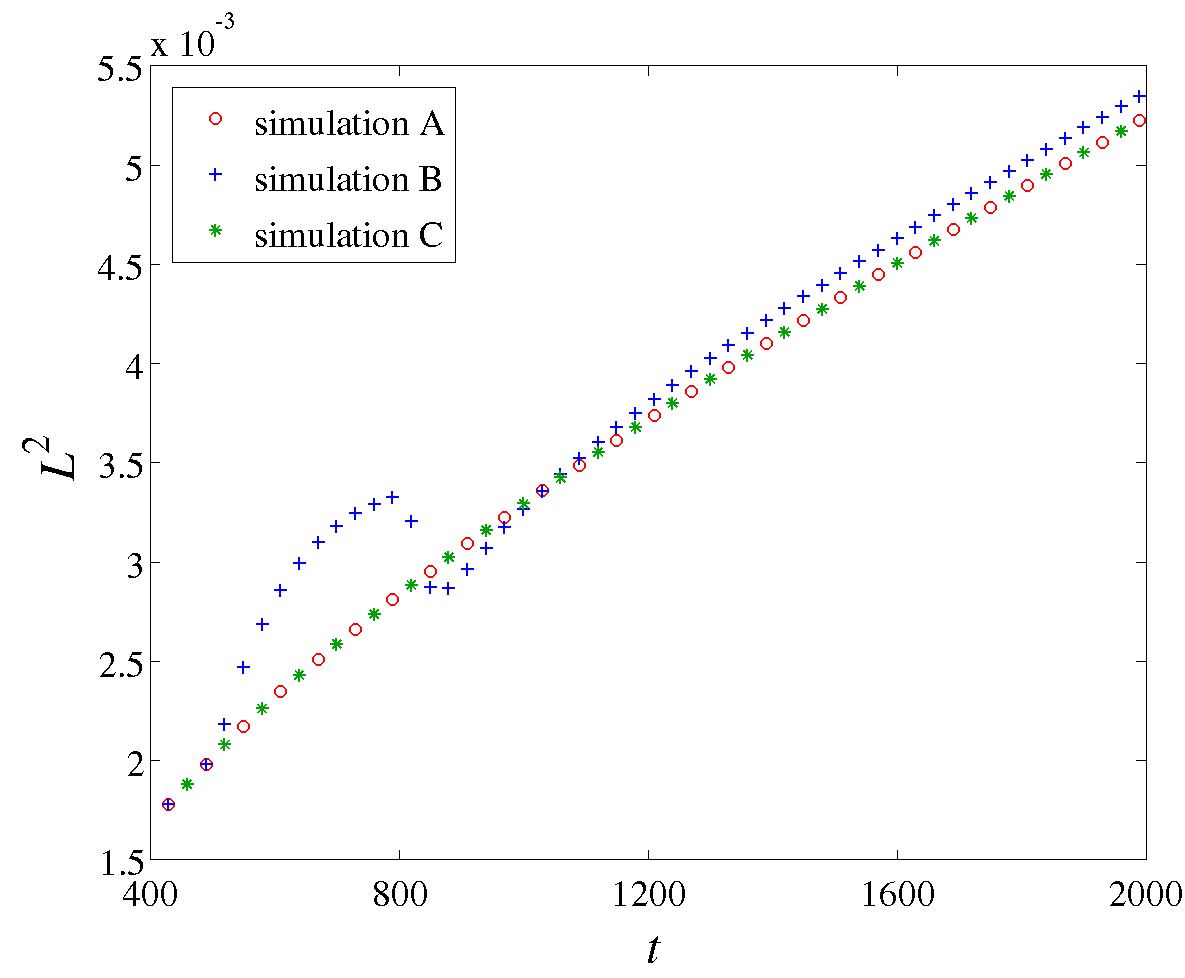}
	\caption{{\small (Color online) Evolution of the characteristic
length, $L(t)$, for simulations $A$, $B$ and $C$. Since the curves for   
simulations $A$ and $C$ are almost indistinguishable we intercalate the
corresponding datapoints in order to improve the visualization.}}
	\label{L(t)} 
\end{figure}

Fig. \ref{S_n-900-B} shows a snapshot of simulation $B$ for
$t= 900$ after the electric field has been rotated by an angle
of $\pi$ and then switched off when at $t=800$. In this case, 
the components of the director are given by ${\bf n}=(n_{x_{0}},-n_{y_{0}},0)$ 
and all the half-integer disclinations are back into the $xy$ plane, with an opposite
topological charge to the one they had in Figure \ref{S_n-500}. A
similar realization of the exchange of sign of all half-integer
disclinations network is possible through the rotation of the electric
field around any other axis of the simulation plane. Besides this
transformation of the topological charge, the electric field produces 
other effects on the network coarsening that need to be addressed 
in more detail.

In simulation $C$ we have implemented by hand the modification of the director and 
co-director profiles corresponding to a rotation of $\pi$ of an
external electric field. This transformation was performed with
the purpose of isolating the effect of the inversion of the sign of the 
half-integer disclinations from the modification of the coarsening 
dynamics associated with the presence of an external electric field. 
We have modified the components of ${\bf n}$ and ${\bf l}$ of 
the configuration at $t= 500$, shown in Fig. \ref{S_n-500}, so that
\begin{eqnarray} 
	{n}_{x} & = &  n_{x0}\,,\qquad  {n}_{y}  =   -n_{y0}\,, \nonumber \\ 
	{l}_{x} & = &  n_{y0}\,,\qquad  {l}_{y}  =   n_{x0}\,,
	\label{changedir}
\end{eqnarray}
where ($n_{x_{0}}$, $n_{y_{0}}$, 0) are the components of the director
${\bf n}$ presented in Fig. \ref{S_n-500}. The results for the
projection of the director onto the lattice plane are presented in
Fig. \ref{S_n-900-C} along with the order parameter $S$ for a snapshot of 
simulation $C$ at $t= 900$. Fig. \ref{S_n-900-C} is very similar to  Fig. \ref{S_n-500}, 
except for the inversion of the sign of the topological charges and the coarsening 
which took place between $t=499$ and $t=900$.

\begin{figure}[htb]
	\includegraphics[scale=0.38]{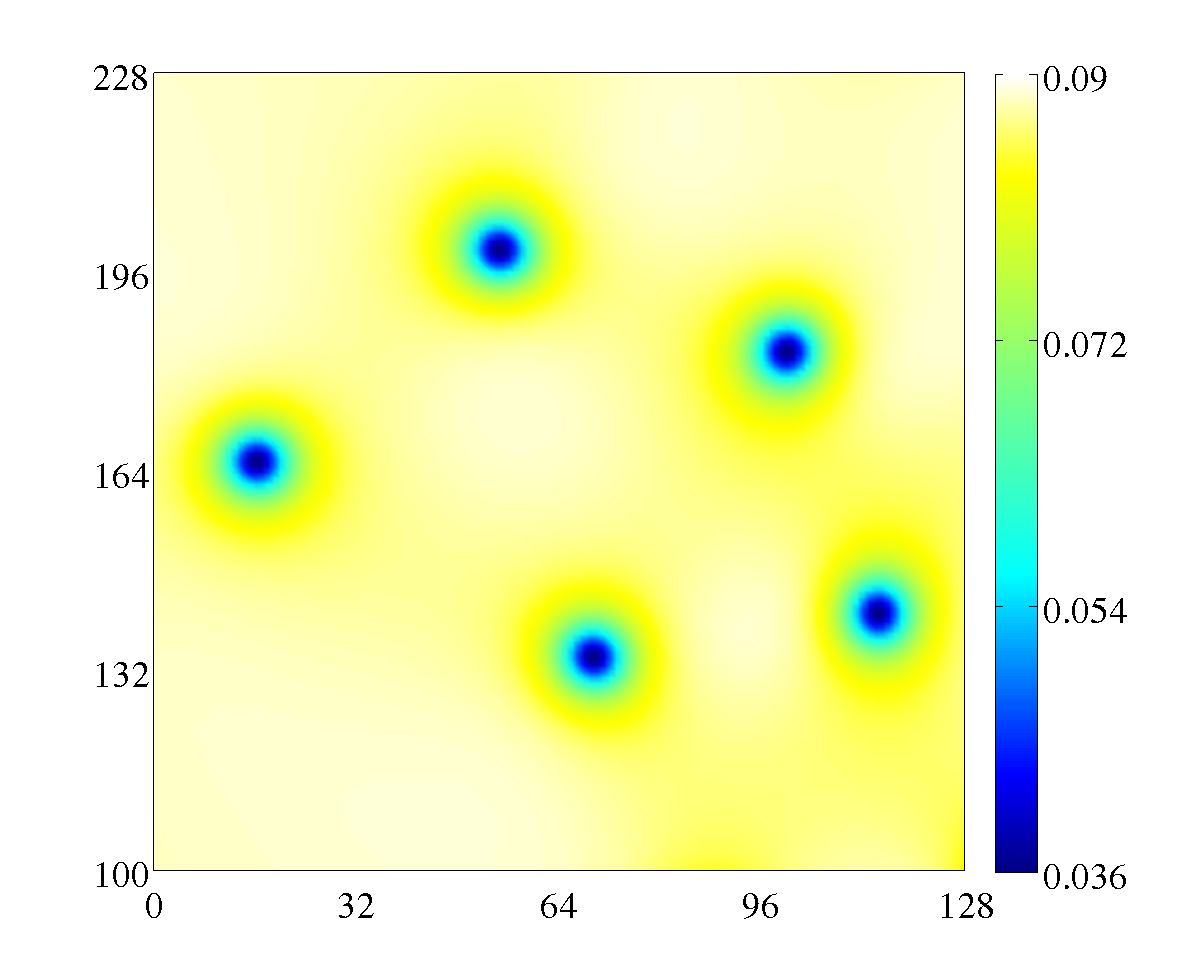}
	\caption{{\small (Color online) Total free energy, given
by Eq. (\ref{free_energy}), for a snapshot of simulation $A$ at $t= 900$. The corresponding 
image for simulation $C$ is identical.}}
\label{E-900} 
\end{figure}

The evolution of the characteristic length, $L(t)$, of simulations $A$, $B$ 
and $C$ is presented in Figure \ref{L(t)}. Given that the curves for   
simulations $A$ and $C$ are almost indistinguishable we intercalate the
corresponding datapoints in order to improve the visualization. The effect of the rotation of the
electric field is visible in simulation $B$, between $t= 500$ and $t=
800$. An acceleration of the coarsening, reflected in the larger slope
of $L^{2}(t)$ may be observed between $t= 500$ ($\varphi=0$) and $t=
650$ ($\varphi=\pi/2$). It is also clear that the memory of the connection 
of the field is preserved by the coarsening
dynamics in simulation $B$, even after the electric field is disconnected. 
This is reflected in the larger value of the characteristic length scale of 
the network at late time, compared to simulations $A$ and $C$.

The comparison of the evolution of $L^{2}(t)$ for simulations $A$ and
$C$ demonstrates that modifying the director and co-director profiles by hand 
using Eq. (\ref{changedir}) does not modify the coarsening dynamics. This happens 
because the bulk and elastic components of the free energy defined by Eqs. (\ref{energy_bulk}) 
and (\ref{energy_elastic}), are invariant by a rotation of the directors around an 
arbitrary axis. Such rotation originates a rotated director and co-director network profile with the
same free energy at every point in space. This result can be confirmed in
Fig. \ref{E-900} which shows the total free energy, given by Eq. (\ref{free_energy}), 
for a snapshot of simulation $A$ at $t= 900$ (the corresponding 
image for simulation $C$ is identical). As expected, the free energy profile is axially 
symmetric around the defect centers. Hence, the acceleration of the defect coarsening 
due to the application of an external electric field is associated with the modification of 
the degree of alignment of the molecules, reflected in the modification of the values of 
the order parameters $S$ and $T$, and it is therefore insensitive to the inversion of the 
sign of the topological charges.

\section{CONCLUSION}
\label{conclusion} 

In this work we have numerically demonstrated that a rotation by $\pi$ of an external electric field around an axis of the plane of a 2D nematic liquid crystal induces an inversion of the sign of all half-integer disclinations, when a negative dielectric anisotropy is considered. We have further analysed the the impact of the exernal electric field on the coarsening dynamics. 

The inversion of the topological charge investigated in the present paper may be realized experimentally in a simple setup. The topological arguments do not depend on the model parameters and apply to the different phases of a liquid crystal. Although we have assumed that the inversion was induced by the rotation of an external electric field, a magnetic field induced transition is also possible in the case of negative diamagnetic anisotropy. 

\section*{ACKNOWLEDGMENTS}
\label{acknowledgments} 

We thank CAPES, CNPq, REDE NANOBIOTEC BRASIL, INCT-FCx (Brazil) and
FCT (Portugal) for partial financial support. 

\bibliography{lc_002}

\begin{thebibliography}{19}
\expandafter\ifx\csname natexlab\endcsname\relax\def\natexlab#1{#1}\fi
\expandafter\ifx\csname bibnamefont\endcsname\relax
  \def\bibnamefont#1{#1}\fi
\expandafter\ifx\csname bibfnamefont\endcsname\relax
  \def\bibfnamefont#1{#1}\fi
\expandafter\ifx\csname citenamefont\endcsname\relax
  \def\citenamefont#1{#1}\fi
\expandafter\ifx\csname url\endcsname\relax
  \def\url#1{\texttt{#1}}\fi
\expandafter\ifx\csname urlprefix\endcsname\relax\def\urlprefix{URL }\fi
\providecommand{\bibinfo}[2]{#2}
\providecommand{\eprint}[2][]{\url{#2}}

\bibitem[{\citenamefont{de~Gennes and Prost}(1995)}]{deGennes-book}
\bibinfo{author}{\bibfnamefont{P.~G.} \bibnamefont{de~Gennes}}
  \bibnamefont{and} \bibinfo{author}{\bibfnamefont{J.}~\bibnamefont{Prost}},
  \emph{\bibinfo{title}{The Physics of Liquid Crystals}}
  (\bibinfo{publisher}{Clarendon Press, Oxford}, \bibinfo{year}{1995}),
  \bibinfo{edition}{2nd} ed.

\bibitem[{\citenamefont{Kleman and Lavrentovich}(2003)}]{Kleman-book}
\bibinfo{author}{\bibfnamefont{M.}~\bibnamefont{Kleman}} \bibnamefont{and}
  \bibinfo{author}{\bibfnamefont{O.~D.} \bibnamefont{Lavrentovich}},
  \emph{\bibinfo{title}{Soft Matter Physics: An Introduction}}
  (\bibinfo{publisher}{Springer}, \bibinfo{year}{2003}).

\bibitem[{\citenamefont{Vilenkin and Shellard}(1994)}]{1994csot.book.....V}
\bibinfo{author}{\bibfnamefont{A.}~\bibnamefont{Vilenkin}} \bibnamefont{and}
  \bibinfo{author}{\bibfnamefont{E.~P.~S.} \bibnamefont{Shellard}},
  \emph{\bibinfo{title}{Cosmic Strings and Other Topological Defects}}
  (\bibinfo{publisher}{Cambridge University Press},
  \bibinfo{address}{Cambridge}, \bibinfo{year}{1994}).

\bibitem[{\citenamefont{Chuang et~al.}(1991)\citenamefont{Chuang, Durrer,
  Turok, and Yurke}}]{Chuang-Science.251.4999}
\bibinfo{author}{\bibfnamefont{I.}~\bibnamefont{Chuang}},
  \bibinfo{author}{\bibfnamefont{R.}~\bibnamefont{Durrer}},
  \bibinfo{author}{\bibfnamefont{N.}~\bibnamefont{Turok}}, \bibnamefont{and}
  \bibinfo{author}{\bibfnamefont{B.}~\bibnamefont{Yurke}},
  \bibinfo{journal}{Science} \textbf{\bibinfo{volume}{251}},
  \bibinfo{pages}{1336} (\bibinfo{year}{1991}), ISSN \bibinfo{issn}{0036-8075}.

\bibitem[{\citenamefont{Chuang et~al.}(1993)\citenamefont{Chuang, Yurke,
  Pargellis, and Turok}}]{Chuang-PhysRevE.47.3343}
\bibinfo{author}{\bibfnamefont{I.}~\bibnamefont{Chuang}},
  \bibinfo{author}{\bibfnamefont{B.}~\bibnamefont{Yurke}},
  \bibinfo{author}{\bibfnamefont{A.~N.} \bibnamefont{Pargellis}},
  \bibnamefont{and} \bibinfo{author}{\bibfnamefont{N.}~\bibnamefont{Turok}},
  \bibinfo{journal}{Phys. Rev. E} \textbf{\bibinfo{volume}{47}},
  \bibinfo{pages}{3343} (\bibinfo{year}{1993}).

\bibitem[{\citenamefont{Zapotocky et~al.}(1995)\citenamefont{Zapotocky,
  Goldbart, and Goldenfeld}}]{Zapotocky-PhysRevE.51.1216}
\bibinfo{author}{\bibfnamefont{M.}~\bibnamefont{Zapotocky}},
  \bibinfo{author}{\bibfnamefont{P.~M.} \bibnamefont{Goldbart}},
  \bibnamefont{and}
  \bibinfo{author}{\bibfnamefont{N.}~\bibnamefont{Goldenfeld}},
  \bibinfo{journal}{Phys. Rev. E} \textbf{\bibinfo{volume}{51}},
  \bibinfo{pages}{1216} (\bibinfo{year}{1995}).

\bibitem[{\citenamefont{Digal et~al.}(1999)\citenamefont{Digal, Ray, and
  Srivastava}}]{Digal-PhysRevLett.83.5030}
\bibinfo{author}{\bibfnamefont{S.}~\bibnamefont{Digal}},
  \bibinfo{author}{\bibfnamefont{R.}~\bibnamefont{Ray}}, \bibnamefont{and}
  \bibinfo{author}{\bibfnamefont{A.~M.} \bibnamefont{Srivastava}},
  \bibinfo{journal}{Phys. Rev. Lett.} \textbf{\bibinfo{volume}{83}},
  \bibinfo{pages}{5030} (\bibinfo{year}{1999}).

\bibitem[{\citenamefont{Denniston et~al.}(2001)\citenamefont{Denniston,
  Orlandini, and Yeomans}}]{Denniston-PhysRevE.64.021701}
\bibinfo{author}{\bibfnamefont{C.}~\bibnamefont{Denniston}},
  \bibinfo{author}{\bibfnamefont{E.}~\bibnamefont{Orlandini}},
  \bibnamefont{and} \bibinfo{author}{\bibfnamefont{J.~M.}
  \bibnamefont{Yeomans}}, \bibinfo{journal}{Phys. Rev. E}
  \textbf{\bibinfo{volume}{64}}, \bibinfo{pages}{021701}
  (\bibinfo{year}{2001}).

\bibitem[{\citenamefont{Dutta and Roy}(2005)}]{Dutta-PhysRevE.71.026119}
\bibinfo{author}{\bibfnamefont{S.}~\bibnamefont{Dutta}} \bibnamefont{and}
  \bibinfo{author}{\bibfnamefont{S.~K.} \bibnamefont{Roy}},
  \bibinfo{journal}{Phys. Rev. E} \textbf{\bibinfo{volume}{71}},
  \bibinfo{pages}{026119} (\bibinfo{year}{2005}).

\bibitem[{\citenamefont{de~L\'ozar et~al.}(2005)\citenamefont{de~L\'ozar,
  Sch\"opf, Rehberg, Sven\ifmmode~\check{s}\else \v{s}\fi{}ek, and
  Kramer}}]{Lozar-PhysRevE.72.051713}
\bibinfo{author}{\bibfnamefont{A.}~\bibnamefont{de~L\'ozar}},
  \bibinfo{author}{\bibfnamefont{W.}~\bibnamefont{Sch\"opf}},
  \bibinfo{author}{\bibfnamefont{I.}~\bibnamefont{Rehberg}},
  \bibinfo{author}{\bibfnamefont{D.}~\bibnamefont{Sven\ifmmode~\check{s}\else
  \v{s}\fi{}ek}}, \bibnamefont{and}
  \bibinfo{author}{\bibfnamefont{L.}~\bibnamefont{Kramer}},
  \bibinfo{journal}{Phys. Rev. E} \textbf{\bibinfo{volume}{72}},
  \bibinfo{pages}{051713} (\bibinfo{year}{2005}).

\bibitem[{\citenamefont{Mukai et~al.}(2007)\citenamefont{Mukai, Fernandes,
  de~Oliveira, and Dias}}]{Mukai-PhysRevE.75.061704}
\bibinfo{author}{\bibfnamefont{H.}~\bibnamefont{Mukai}},
  \bibinfo{author}{\bibfnamefont{P.~R.~G.} \bibnamefont{Fernandes}},
  \bibinfo{author}{\bibfnamefont{B.~F.} \bibnamefont{de~Oliveira}},
  \bibnamefont{and} \bibinfo{author}{\bibfnamefont{G.~S.} \bibnamefont{Dias}},
  \bibinfo{journal}{Phys. Rev. E} \textbf{\bibinfo{volume}{75}},
  \bibinfo{pages}{061704} (\bibinfo{year}{2007}).

\bibitem[{\citenamefont{Bhattacharjee et~al.}(2008)\citenamefont{Bhattacharjee,
  Menon, and Adhikari}}]{Bhattacharjee-PhysRevE.78.026707}
\bibinfo{author}{\bibfnamefont{A.~K.} \bibnamefont{Bhattacharjee}},
  \bibinfo{author}{\bibfnamefont{G.~I.} \bibnamefont{Menon}}, \bibnamefont{and}
  \bibinfo{author}{\bibfnamefont{R.}~\bibnamefont{Adhikari}},
  \bibinfo{journal}{Phys. Rev. E} \textbf{\bibinfo{volume}{78}},
  \bibinfo{pages}{026707} (\bibinfo{year}{2008}).

\bibitem[{\citenamefont{Kitzerow}(1991)}]{Kitzerow-MolCrystLiqCryst.202.51}
\bibinfo{author}{\bibfnamefont{H.~S.} \bibnamefont{Kitzerow}},
  \bibinfo{journal}{Mol. Cryst. Liq. Cryst.} \textbf{\bibinfo{volume}{202}},
  \bibinfo{pages}{51} (\bibinfo{year}{1991}).

\bibitem[{\citenamefont{Dierking et~al.}(2005)\citenamefont{Dierking, Marshall,
  Wright, and Bulleid}}]{Dierking-PhysRevE.71.061709}
\bibinfo{author}{\bibfnamefont{I.}~\bibnamefont{Dierking}},
  \bibinfo{author}{\bibfnamefont{O.}~\bibnamefont{Marshall}},
  \bibinfo{author}{\bibfnamefont{J.}~\bibnamefont{Wright}}, \bibnamefont{and}
  \bibinfo{author}{\bibfnamefont{N.}~\bibnamefont{Bulleid}},
  \bibinfo{journal}{Phys. Rev. E} \textbf{\bibinfo{volume}{71}},
  \bibinfo{pages}{061709} (\bibinfo{year}{2005}).

\bibitem[{\citenamefont{Alexander and
  Marenduzzo}(2008)}]{Alexander-EPL.81.66004}
\bibinfo{author}{\bibfnamefont{G.~P.} \bibnamefont{Alexander}}
  \bibnamefont{and}
  \bibinfo{author}{\bibfnamefont{D.}~\bibnamefont{Marenduzzo}},
  \bibinfo{journal}{EPL} \textbf{\bibinfo{volume}{81}}, \bibinfo{pages}{66004}
  (\bibinfo{year}{2008}).

\bibitem[{\citenamefont{Fukuda et~al.}(2009)\citenamefont{Fukuda, Yoneya, and
  Yokoyama}}]{Fukuda-PhysRevE.80.031706}
\bibinfo{author}{\bibfnamefont{J.-i.} \bibnamefont{Fukuda}},
  \bibinfo{author}{\bibfnamefont{M.}~\bibnamefont{Yoneya}}, \bibnamefont{and}
  \bibinfo{author}{\bibfnamefont{H.}~\bibnamefont{Yokoyama}},
  \bibinfo{journal}{Phys. Rev. E} \textbf{\bibinfo{volume}{80}},
  \bibinfo{pages}{031706} (\bibinfo{year}{2009}).

\bibitem[{\citenamefont{Fukuda}(2010)}]{Fukuda-PhysRevE.81.040701}
\bibinfo{author}{\bibfnamefont{J.-i.} \bibnamefont{Fukuda}},
  \bibinfo{journal}{Phys. Rev. E} \textbf{\bibinfo{volume}{81}},
  \bibinfo{pages}{040701} (\bibinfo{year}{2010}).

\bibitem[{\citenamefont{de~Oliveira et~al.}(2010)\citenamefont{de~Oliveira,
  Avelino, Moraes, and Oliveira}}]{deOliveira-PhysRevE.82.041707}
\bibinfo{author}{\bibfnamefont{B.~F.} \bibnamefont{de~Oliveira}},
  \bibinfo{author}{\bibfnamefont{P.~P.} \bibnamefont{Avelino}},
  \bibinfo{author}{\bibfnamefont{F.}~\bibnamefont{Moraes}}, \bibnamefont{and}
  \bibinfo{author}{\bibfnamefont{J.~C. R.~E.} \bibnamefont{Oliveira}},
  \bibinfo{journal}{Phys. Rev. E} \textbf{\bibinfo{volume}{82}},
  \bibinfo{pages}{041707} (\bibinfo{year}{2010}).

\bibitem[{\citenamefont{de~Gennes}(1971)}]{deGennes-MolCrystLiqCryst.12.193}
\bibinfo{author}{\bibfnamefont{P.~G.} \bibnamefont{de~Gennes}},
  \bibinfo{journal}{Mol. Cryst. Liq. Cryst.} \textbf{\bibinfo{volume}{12}},
  \bibinfo{pages}{193} (\bibinfo{year}{1971}).

\end{thebibliography}


\end{document}